\documentclass[12pt,article]{article}
\usepackage[T2A]{fontenc}
\usepackage[utf8]{inputenc}
\usepackage{amsthm,amsfonts,amssymb,amscd,cite}
\usepackage[english]{babel}
\usepackage[pdftex]{graphicx}
\usepackage{graphicx}
\usepackage[tbtags]{amsmath}
\usepackage{hyperref}

\DeclareSymbolFont{bletters}{OML}{cmm}{bx}{it}
\DeclareMathSymbol{\bla}{\mathord}{bletters}{'025}
\DeclareMathSymbol{\bmu}{\mathord}{bletters}{'026}
\DeclareMathSymbol{\bth}{\mathord}{bletters}{'022}
\DeclareMathSymbol{\bfI}{\mathord}{bletters}{"49}
\DeclareMathSymbol{\bdl}{\mathord}{bletters}{"0E}
\DeclareMathSymbol{\bDl}{\mathord}{bletters}{"001}

\def \bmu{\boldsymbol\mu}
\def \bla{\boldsymbol\lambda}

\begin{document}
\title {
$${}$$\\
{\bf The ground state-vector of the $XY$ Heisenberg chain and the Gauss decomposition}}
\author{
{\bf Nikolay Bogoliubov, Cyril Malyshev}\\
$${}$$\\
{\it St.-Petersburg Department of Steklov Institute of Mathematics, RAS}\\
{\it Fontanka 27, St.-Petersburg,
RUSSIA}
}

\date{}

\maketitle

\def \Ga{\Gamma}
\def \ga{\gamma}
\def \si{\sigma}
\def \la{\lambda}
\def \be{\beta}
\def \al{\alpha}
\def \ta{\theta}
\def \dl{\delta}
\def \Om{\Omega}
\def \Dl{\Delta}
\def \CA{\mathcal A}
\def \CK{\mathcal M}
\def \CM{\mathcal M}
\def \CN{\mathcal N}
\def \cN{\mathcal N}
\def \CV{\mathcal V}
\def \CU{\mathcal U}
\def \CO{\mathcal O}
\def \CC{\mathcal C}
\def \CH{\mathcal H}
\def \CP{\mathcal P}
\def \CZ{\mathcal Z}
\def \CT{\mathcal T}
\def \BC{\mathbb{C}}
\def \BN{\mathbb{N}}
\def \BR{\mathbb{R}}
\def \BZ{\mathbb{Z}}
\def \BI{\mathbb{I}}
\def \inf{{\rm{SA}}}
\def \d{{\rm{d}}}
\def \w{\widetilde}
\def \h{\widehat}
\def \ep{\epsilon}
\def \diag{{\rm diag}\,}
\def \tr{{\rm tr}\,}

\begin{abstract}

\noindent The $XY$ Heisenberg spin$\frac12$ chain is considered in the fermion representation. The construction of the ground state-vector is based on the group-theoretical approach. The exact expression for the ground state-vector will allow to study the combinatorics of the correlation functions of the model.
\end{abstract}


\thispagestyle{empty}
\newpage

\section{Introduction}
\label{sec1}

The correlation functions of certain quantum integrable models demonstrate connection with enumerative combinatorics and with the theory of symmetric functions \cite{fran, bmumn, ramis, nest, statm}. For instance, random lattice walks and boxed plane partitions, as subjects of enumerative combinatorics \cite{stan1}, are related to
the correlation functions of the $XX$ model \cite{b1, b11, b2, b3, bmnph}. Various spin lattice models \cite{fred}, including the $XY$ Heisenberg chain model, as well as its isotropic limit, the $XX$ model, provide a base for such actively developing subjects in the theory of quantum information \cite{vk2} as random lattice walks \cite{krattmon1} and
entanglement entropy \cite{sugi}.

\begin{figure}[h]
\center
\includegraphics {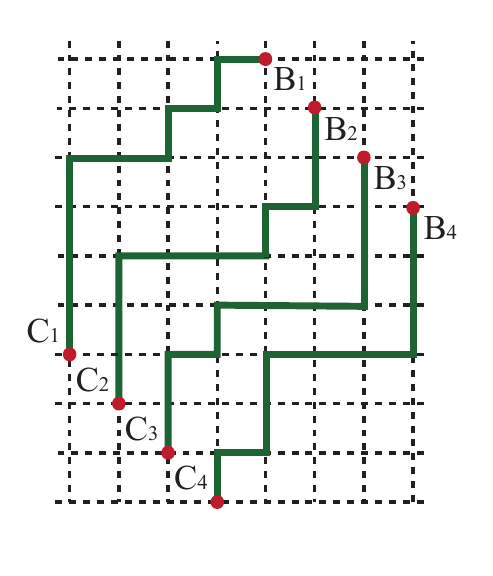}
\caption{A nest of self-avoiding lattice paths called \textit{watermelon}}
\label{fig:f44}
\end{figure}

Interest in the study of the correlation functions for the $XY$ spin chain still exists after the pioneer works \cite{lb, nj, mc1}.
The determinantal representation for the the equal-time correlation functions for the model was obtained in the paper \cite{iz1}.
The approach based on the application of the coherent states to the problem of time and temperature dependent correlation functions was developed in
\cite{iz2, pron}. In the present paper we consider a group theoretical approach to the $XY$ model and derive its ground state-vector using the \textit{Gauss decomposition} \cite{perel}.
The representation of the obtained ground state wave function  would be of importance in the combinatorial interpretation of the correlation functions of the model. Namely, the scalar products of the state-vectors may be described as a linear combination of nests of self-avoiding lattice paths, so-called, \textit{watermelons} (see Figure \ref{fig:f44} and Ref.~\cite{bmumn}).

The paper is organised as follows.
Section~\ref{sec1} is introductory. The fermion representation of the model is provided in Section~\ref{sec2}, and diagonalization of the Hamiltonian by means of the Bogoliubov transformation is performed. The
ground state wave function is derived in Section~\ref{sec3}.
Section~\ref{sec4} concludes the paper.

\section{Outline of the model}
\label{sec2}

The $XY$ Heisenberg spin chain is
described by the Hamiltonian, \cite{lb, nj}:
\begin{align}
H=H_{\rm xx}+\gamma H_{\rm anis}\,,\qquad H_{\rm xx}\equiv-\displaystyle{\sum \limits^{M}_{n, m=1}} \Dl^{(+)}_{nm}\si^+_{n}\si^-_{m}\,, & \label{cor:lin2}
\\
H_{\rm anis} \equiv-\displaystyle{\frac12} \sum\limits_{n, m=1}^M \Dl^{(+)}_{n m}
(\si^+_n\si^+_{m}+\si^-_n \si^-_{m})\,, & \label{cor:lin3}
\end{align}
where $\ga$
is anisotropy parameter, and $M=0\pmod{2}$ is the number of sites. The local spin operators
$\si^\pm_n=\frac12 (\si^x_n \pm i\si^y_n)$ and $\si^z_n$ depend on the lattice argument
$n\in\mathcal M \equiv \{1, 2, \dots, M\}$ and satisfy the commutation relations:
\begin{equation}
\label{cor:lin222}
[\si^+_k, \si^-_l]\,=\,\dl_{k l} \si^z_l\,,\qquad
[\si^z_k,\si^\pm_l]\,=\,\pm2 \dl_{k l} \si^{\pm}_l\,.
\end{equation}
The entries of the \textit{hopping matrix} $\Dl^{(s)}$ in (\ref{cor:lin2}), (\ref{cor:lin3}) are expressed as follows, \cite{b2}:
\begin{equation}
\Dl^{(s)}_{nm}\equiv\frac12\,
(\dl_{|n-m|, 1}+s\dl_{|n-m|, M-1}),
\label{cor:lin4}
\end{equation}
where ${\dl}_{n, l} (\equiv {\dl}_{n l})$ is the Kronecker symbol, and $s$ is either $\pm 1$ or zero. The periodic boundary conditions
$\si^{\al}_{n+M}=\si^{\al}_n$, $\forall n\in\mathcal M$, are imposed. The Hamiltonian $H_{\rm xx}$ (\ref{cor:lin2}) (i.e., $H$ at $\ga=0$) is that of the periodic $XX$ chain.

Let us pass from the spin operators $\sigma^\al_k$
to the canonical fermion operators $c_k$, $c^\dagger_k$ subjected to the algebra \begin{equation}
\label{cor:lin7101}
\{c_k, c_n\} = \{c^\dagger_k, c^\dagger_n\}=0, \quad
\{c_k, c^\dagger_n\}=\dl_{kn}\,,
\end{equation}
where the brackets $\{\,,\,\}$ imply anti-commutation. We use the Jordan-Wigner transformation \cite{jor}:
\begin{equation}
c_k = \exp\Big(
i\pi\sum \limits^{k-1}_{n=1} \sigma^-_n\sigma^+_n\Big)
\,\sigma^+_k, \qquad
c^\dagger_k = \sigma^-_k\,\exp\Big(
\!-i\pi\sum\limits^{k-1}_{n=1} \sigma^-_n\sigma^+_n\Big)\,.
\label{cor:lin7}
\end{equation}
Inversion of (\ref{cor:lin7})
takes the form:
\begin{equation}
\label{cor:lin731}
\si^-_n\,=\,c_n^\dagger \prod\limits_{j=1}^{n-1} (1-2 c_j^\dagger c_j)\,, \qquad \si^+_n\,=\, \prod\limits_{j=1}^{n-1} (1-2 c_j^\dagger c_j)\,c_n\,.
\end{equation}
The periodic boundary
conditions for the spin variables are equivalent to the following boundary conditions for the fermion variables:
\begin{equation}
c_{M+1}=(-1)^{\mathcal{N}}c_1, \qquad c^{\dagger}_{M+1} = c^{\dagger}_1 (-1)^{\mathcal{N}}\,,
\label{cor:lin8}
\end{equation}
where ${\cN} \equiv Q(M)=\sum^M_{k=1}
c^\dagger_k c_k$ is the total number of particles.

The transformations (\ref{cor:lin7}), (\ref{cor:lin731}) enable to represent $H$
(\ref{cor:lin2}) as follows \cite{lb, nj}:
\begin{align}
H\,=& \,H^+P^+ + H^-P^-\,,\label{cor:lin9} \\
H^\pm = & -\frac 12\sum\limits^M_{k=1} \big[c^\dagger_k c_{k+1}
+c^\dagger_{k+1} c_k + \gamma(c_{k+1}c_k +
c^\dagger_kc^\dagger_{k+1})\big]\,,
\label{cor:lin10}
\end{align}
where $P^\pm$ are the projectors onto the states with even ($+$)/odd ($-$) number of fermions: $P^\pm$ $=$ $\frac12 (1\pm(-1)^{\cN})$. The indices $s=\pm$ point out the correspondence between the operators $H^s$ (\ref{cor:lin10})
and appropriately specified boundary conditions (\ref{cor:lin8}):
\begin{equation}\label{cor:lin811}
c_{M+1} = -s c_1, \qquad c^\dagger_{M+1} =
-s c^\dagger_1\,.
\end{equation}
The operator ${\cN}$ commutes with $H_{\rm xx}$ (\ref{cor:lin2}). The parity operator
$(-1)^{{\cN}}$ commutes with $H$ and anti-commutes with $c^\dagger_k$ and $c_k$.

The requirements
(\ref{cor:lin811}) suggest to use the Fourier series:
\begin{equation}
c_k = \frac{e^{-i\pi/4}}{\sqrt M} \sum\limits_{q\in {\sf S}^\pm}
e^{i q k} c_q\,,\qquad
c^{\dagger}_k = \frac{e^{i\pi/4}}{\sqrt M} \sum\limits_{q\in {\sf S}^\pm}
e^{-i q k} c^{\dagger}_q\,,
\label{cor:lin22}
\end{equation}
where $\sum_{q\in {\sf S}^\pm}$
implies summation over
quasi-momenta $q \in {\sf S}^\pm$ respecting $\cos Mq = \mp 1$:
\begin{align}
\nonumber
&{\sf S}^+ =\{q:q=-\pi+\pi(2l-1)/M
\,,\,\,l \in \CM\}\,,\\[0.2cm]
\nonumber
&{\sf S}^- =\{q:q=-\pi+2\pi l/M\,,\,\,l \in \CM\}\,.
\end{align}
Substitution of (\ref{cor:lin22}) into (\ref{cor:lin10}) yields the $XY$ Hamiltonian in the momentum representation:
\begin{align}
\label{cor:lin25}
H^{\pm} = \frac12 \,\sum_{q\in {\sf S}^\pm} (c^\dagger_q, c_{-q}) {\cal H}_q \begin{pmatrix} c_q \\ c^\dagger_{-q} \end{pmatrix}\,, \\
\label{cor:lin26}
{\cal H}_q = \ep_q\,\si^z + \Ga_q\, (\si^++\si^-)\,,
\end{align}
where $\ep_q\equiv -\cos q$ and $\Ga_q\equiv \ga\sin q$.

It appropriate to introduce three quadratic operators ${\cal J}^{\pm}_q$,
${\cal J}_{q}^{0}$ expressed through the fermion operators $c_q$ and $c^{\dagger}_q$:
\begin{align}
\label{cor:lin27}
& {\cal J}^{-}_q= c_{-q} c_{q}\,, \quad {\cal J}_{q}^{+} = c^{\dagger}_{q} c^{\dagger}_{-q}\,, \\
\label{cor:lin28}
& {\cal J}_{q}^{0} = \frac12 (c^{\dagger}_{q} c_{q} + c^{\dagger}_{-q} c_{-q} - 1)\,.
\end{align}
The operators (\ref{cor:lin27}),  (\ref{cor:lin28}) are related to the algebra $\mathfrak{su}(2)$
since satisfy the
commutation relations of the form (compare with (\ref{cor:lin222})):
\begin{equation}
\label{cor:lin31}
[{\cal J}^{+}_q, {\cal J}_{p}^{-} ] = 2 {\cal J}^{0}_q \dl_{p q}\,, \qquad [{\cal J}^{0}_q, {\cal J}_{p}^{\pm} ] = \pm {\cal J}^{\pm}_q \dl_{p q}\,.
\end{equation}
The definitions (\ref{cor:lin27}) and (\ref{cor:lin28}) allow us to express $H^{\pm}$ (\ref{cor:lin25}) as follows:
\begin{equation}
\label{cor:lin33}
H^{\pm} = \sum_{q\in {\sf S}^\pm} \ep_q\,{\cal J}^{0}_q + \frac{\Ga_q}2\, ({\cal J}^{+}_q + {\cal J}_{q}^{-})\,.
\end{equation}

Let us relate the canonical operators $c^{\dagger}_q$, $c_q$ to the new fermionic operators $A^{\dagger}_q$, $A_q$ by means
of the unitary matrix $g^\dagger_{\theta} \in SU(2, \BR)$,
\begin{align}
\label{cor:lin34}
& \begin{pmatrix} c_q \\ c^\dagger_{-q} \end{pmatrix} = g^{\dagger}_{\theta} \begin{pmatrix} A_q \\ A^\dagger_{-q} \end{pmatrix}\,, \\
\label{cor:lin35}
& g^{\dagger}_{\theta} =
e^{-{\theta_q}
(\si^--\si^+)} = e^{i {\theta_q} \si^y} =
\cos {\theta_q}\,\si^0 + \sin {\theta_q} (\si^+-\si^-)\,.
\end{align}
The transformation (\ref{cor:lin34}) and its conjugated are used in  (\ref{cor:lin25}), and it enables to diagonalize the matrix ${\cal H}_q$ (\ref{cor:lin26})
as follows:
\begin{equation}
\label{cor:lin36}
g_{\theta}\, {\cal H}_q \, g^{\dagger}_{\theta} = E_q \, \si^z\,.
\end{equation}
The relation (\ref{cor:lin36})
is equivalent to the following equations:
\begin{align}
\label{cor:lin37}
& \ep_q
\cos 2{\theta_q} - \Ga_q \sin 2{\theta_q} = E_q \,,  \\[0.1cm]
\label{cor:lin38}
& \ep_q \sin 2{\theta_q} + \Ga_q \cos 2{\theta_q} = 0 \,,
\end{align}
where $E_q =(\ep_q^2 + \Ga^2_q)^{1/2}$. It follows from (\ref{cor:lin38}) that $\theta_q$ respects $\tan 2\theta_q = - \Ga_q/\ep_q$.

Let us introduce, analogously to (\ref{cor:lin27}), (\ref{cor:lin28}), the appropriate operators $\bar{\cal J}^{\pm}_q$,
$\bar{\cal J}_{q}^{0}$ in terms of the fermion operators $A_q$ and $A^{\dagger}_q$:
\begin{align}
\label{cor:lin39}
& {\bar{\cal J}}^{-}_q= A_{-q} A_{q}\,, \quad {\bar{\cal J}}_{q}^{+} = A^{\dagger}_{q} A^{\dagger}_{-q}\,, \\
\label{cor:lin40}
& {\bar{\cal J}}_{q}^{0} = \frac12 (A^{\dagger}_{q} A_{q} + A^{\dagger}_{-q} A_{-q} - 1)\,.
\end{align}
The following transformation  takes place:
\begin{align}
\label{cor:lin41}
{{\cal J}}^{-}_q +
{{\cal J}}^{+}_q = ({\bar{\cal J}}^{-}_q +
{\bar{\cal J}}^{+}_q) \cos 2{\theta_q}
- 2 {\bar{\cal J}}_{q}^{0}
\sin 2{\theta_q}\,, & \\
\label{cor:lin42}
2 {{\cal J}}_{q}^{0} = ({\bar{\cal J}}^{-}_q +
{\bar{\cal J}}^{+}_q) \sin 2{\theta_q}
+ 2 {\bar{\cal J}}_{q}^{0}
\cos 2{\theta_q}\,. &
\end{align}
Applying the transformations (\ref{cor:lin41}) and (\ref{cor:lin42}) allows us to express the Hamiltonians $H^{\pm}$ (\ref{cor:lin33}) as follows:
\begin{equation}
\label{cor:lin43}
H^{\pm} = \sum_{q\in {\sf S}^\pm} E_q\,{\bar{\cal J}}^{0}_q
= \sum_{q\in {\sf S}^\pm} E_q A^{\dagger}_{q} A_{q} + E^{\pm}_{\rm {gr}} \,,
\end{equation}
provided that the relations (\ref{cor:lin37}) and (\ref{cor:lin38}) hold, and $E^{\pm}_{\rm {gr}}$ is expressed as
\begin{equation}
\nonumber
E^{\pm}_{\rm {gr}} = \frac{-1}{2} \sum_{q\in {\sf S}^\pm} E_q\,.
\end{equation}

\section{The ground state wave function}
\label{sec3}

The canonical operators $c_k$, $c^\dagger_k$, as well as $c_q$, $c^\dagger_q$, characterized by the relations (\ref{cor:lin7101})
possess the Fock vacuum $|0\rangle$ (and its conjugate $\langle 0|$):
\begin{align}
\label{cor:lin44}
& c_k |0\rangle =0\,, \quad \langle 0| c^\dagger_k = 0\,,
\qquad k\in {\CM}\,, \\
\label{cor:lin45}
& c_q |0\rangle =0\,, \quad \langle 0| c^\dagger_q = 0\,,
  \qquad q\in {\sf S}^{\pm}\,.
\end{align}
The vacuum $|0\rangle$ is normalized, $\langle 0|0\rangle = 1$, and $|0\rangle$ is the same for both
Hamiltonians $H^{\pm}$.

The ground-state vector $|0\rangle\rangle$ of the Hamiltonian (\ref{cor:lin43})  have to satisfy the relations:
\begin{equation}
\label{cor:lin46}
A_q |0\rangle\rangle =0\,, \quad \langle\langle 0| A^\dagger_q = 0\,,\qquad q\in {\sf S}^{\pm}\,.
\end{equation}
We introduce the unitary operator ${\cal U}_{\theta}$,
\begin{equation}
\label{cor:lin47}
{\cal U}_{\theta}\,\equiv\,
\exp\Bigl(\sum_{q\in {\sf S}^\pm} {\theta_q}({{\cal J}}^{+}_q -
{{\cal J}}^{-}_q)\Bigr)=\prod_{q\in {\sf S}^\pm}\exp\bigl({{\theta_q}({{\cal J}}^{+}_q -
{{\cal J}}^{-}_q)}\bigr)\,,
\end{equation}
where ${\cal J}^{-}_q$ and ${\cal J}^{+}_q$ are defined by  (\ref{cor:lin27}) and \ref{cor:lin31}) and formulate the following

\vskip0.3cm
\noindent\textbf{Proposition 1\,\,} \textit{The relations} (\ref{cor:lin46}) \textit{take place provided that the state $|0\rangle\rangle$ is defined as:}
\begin{equation}
\label{cor:lin48}
|0\rangle\rangle\,=\,{\cal U}_{\theta/2}\,|0\rangle\,.
\end{equation}

\noindent\textbf{Proof\,\,}
The commutation relations are valid:
\begin{align}
\label{cor:lin49}
& [ c_q, \sum_{p\in {\sf S}^\pm} {\theta_p}
{\cal J}^{+}_p]\,=\,2 \theta_q c^{\dagger}_{-q}\,,\quad
[ c^\dagger_q, \sum_{p\in {\sf S}^\pm} {\theta_p}
{\cal J}^{+}_p]\,=\,0\,, \\
& [ c^{\dagger}_{-q}, \sum_{p\in {\sf S}^\pm} {\theta_p}
{\cal J}^{-}_p]\,=\,2 \theta_q c_{q}\,,\quad
[ c_q, \sum_{p\in {\sf S}^\pm} {\theta_p}
{\cal J}^{-}_p]\,=\,0\,,
\label{cor:lin50}
\end{align}
where  the property  $\theta_{-q} = -\theta_{q}$ is used.
Taking into account (\ref{cor:lin49}) and (\ref{cor:lin50}) we obtain:
\begin{align}
\label{cor:lin51}
& {\cal U}_{\theta/2}\,c_q\,{\cal U}^\dagger_{\theta/2}\,=\,A_q\,, \\
\label{cor:lin52}
& {\cal U}_{\theta/2}\, c^\dagger_{-q}\,{\cal U}^\dagger_{\theta/2}\,=\, A^\dagger_{-q}\,.
\end{align}
The state $|0\rangle\rangle$
(\ref{cor:lin48}) is annihilated by $A_q$ (\ref{cor:lin51})  since
$c_q$ annihilates the Fock vacuum $|0\rangle$, Eq.~(\ref{cor:lin45}), and  ${\cal U}^\dagger_{\theta/2}\, {\cal U}_{\theta/2}=1$.
The introduced ground state-vector  (\ref{cor:lin48}) is normalized to unity,
$\langle\langle 0|0\rangle\rangle =1$.

The alternative derivation of \textbf{Proposition 1} is based on the equivalence of the relations (\ref{cor:lin51}), (\ref{cor:lin52}) and the transformation (\ref{cor:lin34}).
The action of operator $A_q$ on the ground state (\ref{cor:lin48}) may be written as
\begin{equation}\label{cor:lin521}
A_q |0\rangle\rangle\,=\, (\cos{\theta}_q c_q - \sin{\theta}_q c^\dagger_{-q})\, {\cal U}_{\theta/2} |0\rangle\,.
\end{equation}
The commutation relations
\begin{align}
\label{cor:lin522}
c_q\, {\cal U}_{\theta/2} &=\,
{\cal U}_{\theta/2}
(\cos{\theta}_q\, c_q + \sin{\theta}_q\, c^\dagger_{-q})\,, \\
\label{cor:lin523}
c^\dagger_{-q}\, {\cal U}_{\theta/2} &=\,
{\cal U}_{\theta/2}
(-\sin{\theta}_q\, c_q + \cos{\theta}_q\, c^\dagger_{-q})\,,
\end{align}
ensure that
\begin{equation}\label{cor:lin524}
A_q |0\rangle\rangle\,=\,{\cal U}_{\theta/2}\, c_q\,|0\rangle\,=\,0\,.
\end{equation}
$\,\,\Box$

The Gauss decomposition, \cite{perel}, which may be obtained
by means of `infinit\-es\-im\-al method' \cite{kirz}, is valid for the matrix $g_{\theta} \equiv \exp(-i \theta \si^y) = \exp\bigl(\theta (\si^--\si^+)\bigr)$ (\ref{cor:lin35}):
\begin{equation}
\label{cor:lin53}
e^{\theta (\si^--\si^+)}\,=\,
e^{-\tan\theta\,\si^+}\,
e^{{\bar\gamma} \si^z}\,e^{\tan\theta\,\si^-}\,,
\end{equation}
where $e^{\bar\gamma}=1/\cos\theta$. With regard to
(\ref{cor:lin53}) we arrive at the decomposition for the elements of the operator ${\cal U}_{\theta}$ (\ref{cor:lin47}):
\begin{equation}
\label{cor:lin54}
\exp\bigl({{\theta_q}({{\cal J}}^{+}_q -
{{\cal J}}^{-}_q)}\bigr)\,=\,
\exp\bigl(\tan{\theta_q}{{\cal J}}^{+}_q\bigr)\,
\exp\bigl( 2{\bar\gamma}_q
{{\cal J}}^{0}_q\bigr)\,
\exp\bigl(- \tan{\theta_q}
{{\cal J}}^{-}_q\bigr)\,,
\end{equation}
where $e^{{\bar\gamma}_q} = 1/\cos\theta_q$. Equation
(\ref{cor:lin54}) suggests to formulate the following

\vskip0.3cm
\noindent\textbf{Proposition 2\,\,} \textit{Provided that the representation} (\ref{cor:lin54}) \textit{holds, the state} $|0\rangle\rangle$ (\ref{cor:lin48}) \textit{acquires the equivalent representation}:
\begin{equation}
|0\rangle\rangle\,=\,\Bigl(
\prod_{q\in {\sf S}^\pm} \cos^{1/2}\theta_q\Bigr)\,\exp \Bigl(\sum_{q\in {\sf S}^\pm} {\frac{\tan\theta_q}{2}}\,{{\cal J}}^{+}_q\Bigr) |0\rangle\,.
\label{cor:lin55}
\end{equation}

\noindent\textbf{Proof\,\,} First of all, the following note is of importance:
\begin{align}
\label{cor:lin551}
& {\cal U}_{\theta/2}\,=\,\exp \Bigl(\sum_{q\in {\sf S}^\pm} {\frac{\theta_q}{2}}({{\cal J}}^{+}_q -
{{\cal J}}^{-}_q)\Bigr)\,=\,\exp \Bigl(\sum_{q^+} {\theta_q}({{\cal J}}^{+}_q -
{{\cal J}}^{-}_q)\Bigr) \\
&=\,
\exp\Bigl(\sum_{q\in {\sf S}^\pm} {\frac{\tan\theta_q}{2}}{{\cal J}}^{+}_q\Bigr)\,
\exp\Bigl(\sum_{q\in {\sf S}^\pm} {\bar\gamma}_q
{{\cal J}}^{0}_q\Bigr)\,
\exp\Bigl(-\sum_{q\in {\sf S}^\pm} {\frac{\tan\theta_q}{2}}
{{\cal J}}^{-}_q\Bigr)\,,
\label{cor:lin552}
\end{align}
where $\sum_{q^+} \equiv \sum_{\{(q\in {\sf S}^\pm) \cap\, (q\in\BR^+)\}}$, and antisymmetry of ${\theta_q}$, ${{\cal J}}^{+}_q$,
${{\cal J}}^{-}_q$ with respect to the reflection of $q$ is taken into account in (\ref{cor:lin551}). The Gauss decomposition  (\ref{cor:lin54}) is used to pass from (\ref{cor:lin551}) to (\ref{cor:lin552}).

From definitions of operators ${\cal J}^{-}_q$ (\ref{cor:lin27}) and ${{\cal J}}^{0}_q$ (\ref{cor:lin28}) it follows that ${\cal J}^{-}_q |0\rangle = 0$,
while ${{\cal J}}^{0}_q |0\rangle = \frac{-1}{2} |0\rangle$. Therefore,
\begin{equation}
\label{cor:lin56}
\exp\Bigl(\sum_{q\in {\sf S}^\pm} {\frac{\tan\theta_q}{2}}
{{\cal J}}^{-}_q\Bigr) |0\rangle = |0\rangle\,,
\end{equation}
and
\begin{equation}
\label{cor:lin57}
\exp\Bigl(\sum_{q\in {\sf S}^\pm} {\bar\gamma_q}
{\cal J}^{0}_q\Bigr) |0\rangle = \prod_{q\in {\sf S}^\pm} e^{-{\bar\gamma_q}/2}\,=\,
\prod_{q\in {\sf S}^\pm} \cos^{1/2}\theta_q\,.
\end{equation}
Thus, we obtain from (\ref{cor:lin552}), (\ref{cor:lin56}) and  (\ref{cor:lin57}) that (\ref{cor:lin55}) is valid. $\,\,\Box$

The representation (\ref{cor:lin55}) of the ground state $|0\rangle\rangle$
coincides with that proposed in \cite{iz2}:
\begin{equation}
\nonumber
|0\rangle\rangle=N^{-1/2}\Omega^+
|0\rangle\,, \qquad \Omega^+=\exp\Biggl(\sum_{p\in {\sf S}^\pm}  \frac{\tan\theta_p}{2}\, c^{\dagger}_p c^{\dagger}_{-p} \Biggr)\,.
\end{equation}
The normalizing factor $N^{-1/2}$, where $N=\langle 0|\,\Omega\, \Omega^+\, |0\rangle$ was calculated in \cite{iz2} as the integral over the Grassmann coherent states, is equal to
\begin{equation}
\label{cor:lin5712}
N^{-1/2}\, = \, \prod_{p^+}\cos\theta_p\,,
\end{equation}
and coincides with the coefficient in (\ref{cor:lin55}). The statement of \textbf{Proposition 2} clarifies the origin of this coefficient from the group theoretical viewpoint.

For the sake of completeness we shall give the direct proof that  the  state $|0\rangle\rangle$
expressed by (\ref{cor:lin55}) is annihilated by the operator $A_q$.
Really, since $A_q$ is given by (\ref{cor:lin51}), it is enough to show that the state
\begin{equation}
\label{cor:lin58}
{\cal U}^\dagger_{\theta/2}
\,\exp \Bigl(\sum_{q\in {\sf S}^\pm} {\frac{\tan\theta_q}{2}}\,{{\cal J}}^{+}_q\Bigr) |0\rangle
\end{equation}
is annihilated by $c_q$.
The Gauss decomposition (\ref{cor:lin552}) admits the ``antinormal'' from looking as follows (see \cite{perel}):
\begin{equation}
\label{cor:anorm}
{\cal U}_{\theta/2}=  \exp\Bigl(-\sum_{q\in {\sf S}^\pm} {\frac{\tan\theta_q}{2}}{{\cal J}}^{-}_q\Bigr)\,
\exp\Bigl(-\sum_{q\in {\sf S}^\pm} {\bar\gamma}_q
{{\cal J}}^{0}_q\Bigr)\,
\exp\Bigl(\sum_{q\in {\sf S}^\pm} {\frac{\tan\theta_q}{2}}
{{\cal J}}^{+}_q\Bigr)\,.
\end{equation}
Substituting the conjugated form
of (\ref{cor:anorm})
into (\ref{cor:lin58}) we obtain the state annihilated by $c_q$:
\begin{equation}
\nonumber
\Bigl(\prod_{q\in {\sf S}^\pm} \cos^{-1/2}\theta_q\Bigr)\,
\exp\Bigl(\sum_{q^+}  {\tan\theta_q}
{{\cal J}}^{-}_q\Bigr)
|0\rangle\,=\,\Bigl(
\prod_{q\in {\sf S}^\pm} \cos^{-1/2}\theta_q\Bigr)\,
|0\rangle\,.
\end{equation}

Let us turn to the state $|0\rangle\rangle$ (\ref{cor:lin55}) and consider the following representation:
\begin{align}
\nonumber
& \Bigl(
\prod_{q\in {\sf S}^\pm} \cos^{-1/2}\theta_q\Bigr)\,|0\rangle
\rangle \,=\,\exp \Bigl(\sum_{q\in {\sf S}^\pm} \frac{\tan{{\theta_q}}}2\,{{\cal J}}^{+}_q\Bigr)\,|0\rangle \\
\label{cor:lin782}
& =\,|0\rangle + \sum_{n=1}^{M/2}
\frac{1}{n !} \sum_{\{q^+_l\}_{1\le l\le n}} \prod_{l=1}^{n} \bigr(T_{q_l}
{{\cal J}}^{+}_{q_l}\bigl) |0\rangle\,,
\quad T_{q_l} \equiv \tan{{\theta_{q_l}}}\,,
\end{align}
where $\sum_{\{q^+_l\}_{1\le l\le n}}\equiv\sum_{q^+_1,\,q^+_2, \ldots,\,q^+_n}$, and
the sum over $n$ is finite since ${{\cal J}}^{+}_{q}$ squared is zero.
The expression in right-hand side of (\ref{cor:lin782})
is similar to that derived in \cite{iz1} as the ground state wave function of $XY$ chain.
Recall that $\tan 2\theta_q = - \Ga_q/\ep_q$ (see (\ref{cor:lin38})) is known, and therefore $\tan\theta_q$ is found in the form:
\begin{equation}
\label{cor:lin80}
T_{q}\,=\,\frac{\ep_q\pm {\sqrt{\ep_q^2+\Ga^2_q}}}{\Ga_q}\,=\, \frac{\ep_q\pm E_q}{\Ga_q}\,,
\end{equation}
where $T_{q}$ is defined in (\ref{cor:lin782}). The answer (\ref{cor:lin80}) fulfils the quadratic equation as the corresponding parameter does in \cite{iz1}.
With regard to (\ref{cor:lin80}) it can be argued that Eq.~(\ref{cor:lin782}) just coincides (up to irrelevant factor) with the ground state wave function found in \cite{iz1}.

\section{Conclusion}
\label{sec4}

The expression for the ground ground state-vector (\ref{cor:lin48})  of the $XY$ spin chain obtained by the group theoretical approach was written in the form (\ref{cor:lin55})  with the help of Gaussian decomposition. The representation (\ref{cor:lin55}) brought to the form (\ref{cor:lin782}) reveals the connection between state-vectors studied in \cite{iz1} and \cite{iz2}.

The approach discussed in the present paper will allow to spread the combinatorial interpretation of the correlation functions developed in \cite{bmumn} for the $XX$ model on the $XY$ case.

\section*{Acknowledgement}

This work was supported by the Russian Science Foundation (grant no.~18-11-00297).

\end{document}